\documentclass[twocolumn,aps,showpacs,prb,epsf,graphics,psfig]{revtex4}
\usepackage{graphicx}

\begin{document}
\title{
Reactive Molecular Dynamics study on the first steps of DNA-damage by free hydroxyl radicals
}

\author{Ramin M. Abolfath$^{1,2}$, A. C. T. van Duin$^3$, Thomas Brabec$^2$}

\affiliation{
$^1$School of Natural Sciences and Mathematics and Department of Materials Science, University of Texas at Dallas, Richardson, TX 75080 \\
$^2$Physics Department, University of Ottawa, Ottawa, ON, K1N 6N5, Canada \\
$^3$Department of Mechanical and Nuclear Engineering, Pennsylvania State University, PA 16802 \\
}

\date{\today}

\begin{abstract}
We employ a large scale molecular simulation based on bond-order ReaxFF to simulate the chemical reaction and study the damage to a large fragment of DNA-molecule in the solution by ionizing radiation.
We illustrate that the randomly distributed clusters of diatomic OH-radicals that are primary products of megavoltage ionizing radiation in water-based systems are the main source of hydrogen-abstraction as well as formation of carbonyl- and hydroxyl-groups in the sugar-moiety that create holes in the sugar-rings. These holes grow up slowly between DNA-bases and DNA-backbone and the damage collectively propagate to DNA single and double strand break.
\end{abstract}
\maketitle


It is known that megavoltage radiation (X/$\gamma$-rays, $\alpha$-particles, and heavy ions) ionizes the water molecule and creates neutral free-radicals and aqueous electrons~\cite{EricHall:book,Pogozelski1998:CR,Tullius2005:CO,Aydogan2008:RR,Friedland1999:EB,Terrisol1990:RPD,Wilson1994:RPD,Nikjoo1995:book,Semenenko2005:RR}.
In particular OH-radicals with a very short life-time that is reported to be within nano-seconds~\cite{Sies1993:EJB}, are major contributors to the single/double strand breaking of the DNA molecules and the nucleotide-base damage, as 2/3 of environment surrounding DNA molecules in the cell-nucleus is composed of water molecules~\cite{Aydogan2008:RR}.
Various effects of the ionizing radiation on biological systems that ranges from the development of genetic aberrations, carcinogenesis to aging, have attributed to the role of free radicals.

Computational modeling is a valuable tool in understanding the basic mechanisms that underlie DNA damage. Great effort has been devoted to the statistical modeling of the damage sites based on Monte-Carlo (MC) sampling, using empirical reaction rates and radiation scattering cross-sections~\cite{Aydogan2008:RR,Friedland1999:EB,Terrisol1990:RPD,Wilson1994:RPD,Nikjoo1995:book,Semenenko2005:RR}. These models are limited to MC sampling on a static structure of DNA or dynamical models based on molecular-mechanics (MM) and empirical force-fields (FF), e.g. AMBER/CHARMM FF, that are developed for simulation of the non-reactive aspects of bio-molecules.

The reactive aspects and the time evolution of the multi-site DNA damages driven by cascade of chemical reactions, that are beyond MM methods and empirical FF, require the calculation of the potential energies on-the-fly using first-principle quantum mechanical (QM) models.
Recently ab-initio simulations of the hydrogen
abstraction were developed~\cite{Mundy2002:JPC,Close2008:JPC,Abolfath2009:JPC,Gervasio2004:CEJ,Abolfath2011:JCP}.
However realistic modeling of DNA molecule with its environment requires extensive computer resources and is a major draw-back of QM methods. The DFT calculation for hydrogen abstraction in vacuum is limited to the initial damage of a single base~\cite{Mundy2002:JPC,Close2008:JPC,Abolfath2009:JPC} or single sugar-moiety~\cite{Abolfath2011:JCP}.
Despite recent advances in QM/MM methods that allows simulation of larger molecules~\cite{Gervasio2004:CEJ} and inclusion of solvation~\cite{Abolfath2011:JCP}, the real time simulation of DNA-damage remains still elusive.

To address the above considerations regarding to the large scale modeling of DNA-damage, we have studied the evolution of randomly distributed hydroxyl-radicals in small pockets surrounding the DNA-molecule at room  temperature using molecular dynamics simulations where the atomic interactions are described by the reactive force field potential, ReaxFF~\cite{vanDuin2001:JPC}. ReaxFF is a general bond-order dependent potential that provides accurate descriptions of bond breaking and bond formation. Recent simulations on a number of
hydrocarbon-oxygen systems~\cite{vanDuin2001:JPC} and graphene-oxides~\cite{Bagri2010:NC} showed that ReaxFF reliably provides energies, transition states, reaction pathways and reactivity trends in agreement with QM calculations and experiments.

To enable a reactive simulation of solvated DNA we combined the recently developed ReaxFF reactive force field parameters for peptide/water systems~\cite{Rahaman2011:JPC} with the ReaxFF description for organophosphates, as used previously by Zhu et al.
~\cite{Zhu2008:TCA} to investigate the active site of RNA polymerase and by Quenneville et al.~\cite{Quenneville2010:PC} to study dimethyl methylphosphonate (DMMP) reactions in silica. ReaxFF is a bond-order dependent empirical force field method, which includes a polarizable charge function, enabling application to a wide range of materials and accurate reproduction of reaction energies and barriers~\cite{LaBrosse2010:JPC,Chenoweth2008:JPC,Cheung2005:JPC,vanDuin2001:JPC}. The ReaxFF peptide/phosphate descriptions is fully transferable with the aqueous-phase ReaxFF descriptions used by Raymand et al.~\cite{Raymand2010:SS} to study water dissociation reactions on zinc oxide surfaces and by Fogarty et al.~\cite{Fogarty2010:JCP} to study water reactions on silica surfaces, and as such was tested against DFT-data describing proton transfer reactions to solvated hydroxyl species.

To validate ReaxFF  for hydroxyl ion/water interaction and in order to describe the chemistry and physics of OH- in the aqueous environment we compared the ReaxFF parameters with a set of DFT-data. The DFT-data was obtained at the X3LYP/6-311G** level~\cite{Bryantsev2009:JCTC,Xu2004:PNAS} and describes two cases (1) binding energies of HO[H$_2$O]$_n$-clusters, with $n=1$ and (2) proton migration between HO-/H$_2$O at HO---OH$_2$ distances ranging from 2.4 to 3.4 $\AA$. While ReaxFF overpredicts the OH-[H$_2$O]$_n$-binding energies for n=1 and n=2, it gives an accurate description for the $n=3$ and $n=4$ cases, which are the most relevant for normal-density aqueous phases. ReaxFF displays excellent agreement for the proton migration barriers for O---O distances of 2.3, 2.48 (global minimum) and 2.8 $\AA$, which are the most relevant cases for normal-density aqueous phases, but overpredicts the barriers for large O---O distances. We will provide a more elaborate description of the ReaxFF water/H$_3$O$^+$/OH$^-$ development in a subsequent publication. The equilibration of the system of DNA-water was performed in solvated 1276-atom DNA-strand with 2500 water molecules at T=300K for 30 ps. We found that during this time-frame the DNA retained its overall helical configuration, indicating that ReaxFF retains the overall structural integrity of the DNA over such time-frames and that reactive events observed during exposure to OH radicals can indeed be associated with the radical reactivity.

We examined the stability of DNA surrounded by water-molecules against ReaxFF by equilibrating a solvated 1276-atom DNA-strand (with 2500 water molecules) at room temperature for 30 ps. We found that during this time-frame the DNA retained its overall helical configuration, indicating that ReaxFF retains the overall structural integrity of the DNA over such time-frames and that reactive events observed during exposure to OH radicals can indeed be associated with the radical reactivity.

In this work we demonstrate that ReaxFF gives valuable insight on the details of the microscopic events observed in agreement with the ab-initio CPMD and QMMM CPMD-GROMACS calculation including:
(1) the DNA-backbone hydrogen abstraction and formation of water molecules
(2) nucleotide-sugar-moiety bond breaking and formation of carbonyl- and hydroxyl-groups in sugar-moiety-rings
(3) nucleotide-nucleotide hydrogen bond disruption,
(4) nucleotide structural damage,
(5) efficiency of the radiation generated OH radicals in making direct hydrogen-abstraction and
(6) deactivation mechanisms of OH-radicals within a cluster of radicals due to formation of ozone-molecules and network of hydrogen bonds.

Most significantly we illustrate that the collective damage that breaks the bonds between nucleotide-bases and DNA-backbone can be attributed to the formation of the carbonyl- and hydroxyl-groups in the sugar-moiety rings. The holes created by OH-radicals between the nucleotide-bases and DNA-backbone grow up and evolve to large holes that contain number of bases and a large segment of DNA-backbone. It further propagates to the structural base-damage, DNA single- and double-strand break.

\begin{figure}
\begin{center}
\includegraphics[width=1.0\linewidth]{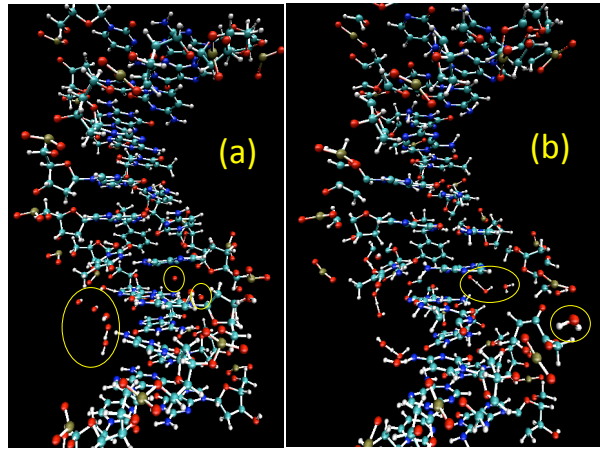}\\ 
\noindent
\caption{
(a) Initial structure of DNA surrounded by randomly generated small pockets of OH-radicals shown in the circles.
(b) DNA structure after $t=0.24$ ps. A large scale double strand damage is clearly visible. Water molecules that are the product of hydrogen-abstraction and two OH-OH bounded by hydrogen bonds are shown in the circles.
Carbon, oxygen, nitrogen, phosphorus and hydrogen atoms are shown as green, red, blue, gold and white, respectively.
}
\label{Fig1}
\end{center}\vspace{-0.5cm}
\end{figure}

\begin{figure}
\begin{center}
\includegraphics[width=1.0\linewidth]{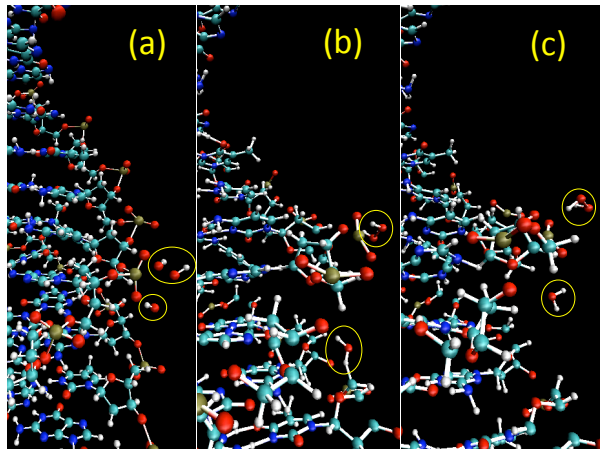}\\ 
\noindent
\caption{
(a) Small cluster of OH radicals close to DNA-backbone is shown in circles. (b) One OH interacts with DNA-backbone and the others combine in form of ozone molecule $H_2O_2$. (c) Hydrogen abstraction from backbone is complete and $H_2O$ molecule forms.
Carbon, oxygen, nitrogen, phosphorus and hydrogen atoms are shown as green, red, blue, gold and white, respectively.
}
\label{Fig2}
\end{center}\vspace{-0.5cm}
\end{figure}

\begin{figure}
\begin{center}
\includegraphics[width=1.0\linewidth]{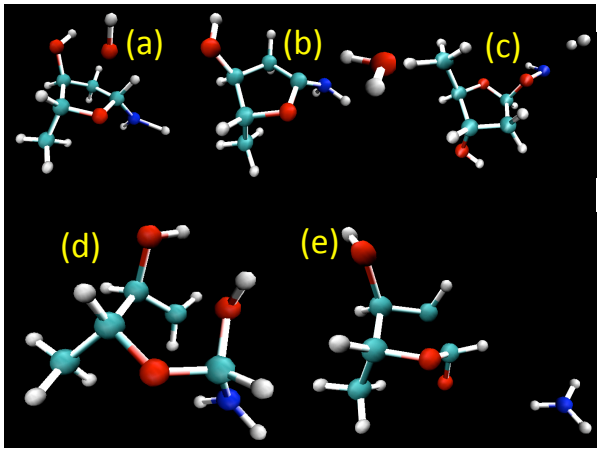}\\ 
\noindent
\caption{
A DFT calculation used for the confirmation of the chemical pathways obtained by ReaxFF. Initial configuration of OH-radical with deoxyribose sugar-moiety attached to amino group (representing the nucleotide-base) shown in (a) leads to formation of water molecule (b), hydrogen molecule (c), hydroxyl group (d), and carbonyl group (e) as a function of initial location of the OH-radical with respect to sugar-moiety ring.
The bond strain created by hydroxyl and carbonyl groups lead to the creation of the hole in the deoxyribose-ring.
Carbon, oxygen, nitrogen and hydrogen atoms are shown as green, red, blue and white, respectively.
}
\label{Fig3}
\end{center}\vspace{-0.5cm}
\end{figure}

\begin{figure}
\begin{center}
\includegraphics[width=1.0\linewidth]{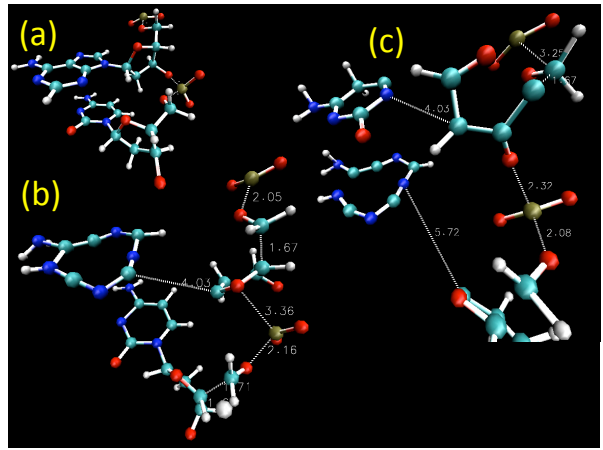}\\ 
\noindent
\caption{
(a) Initial configuration of an adenine in a double strand DNA in the presence of OH-radicals (not shown in the figure) and (b) the configuration of the distorted adenine with a hole (formed by C-C broken bond) at 500fs. The numbers represent the bond-length. The large separation of adenine from the backbone and the stretched backbone that are  indication of the base-damage and single-strand break are clearly visible. For comparison an undamaged cytosine is shown below the adenine.
(c) The damaged cytosine, located one base above the adenine shown in (a) at $t=3500$ fs, illustrating a collective damage to the base, deoxyribose-ring and DNA strand. The opened gap in the deoxyribose pentagon-ring corresponding to C$_{4'}$-O$_{4'}$ missing bond is visible. It indicates formation of stable carbonyl group.
Carbon, oxygen, nitrogen, phosphorus and hydrogen atoms are shown as green, red, blue, gold and white, respectively.
}
\label{Fig4}
\end{center}\vspace{-0.5cm}
\end{figure}

\begin{figure}
\begin{center}
\includegraphics[width=1.0\linewidth]{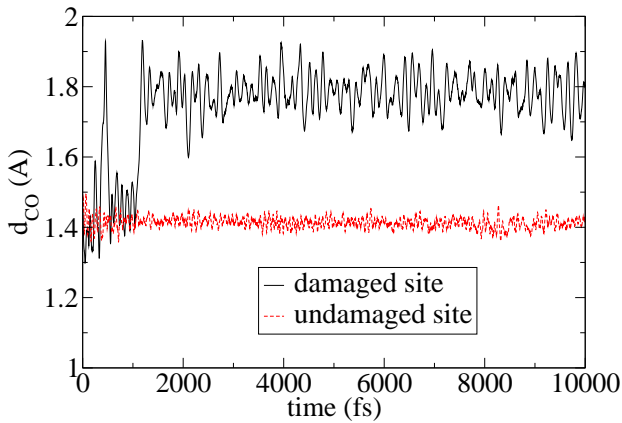} 
\noindent
\caption{
Time-evolution of C$_{4'}$-O$_{4'}$ bond distance, corresponding to the opened gap in the pentagon-ring of deoxyribose shown in Fig.~\ref{Fig4}(c). As a result of series of chemical reactions due to OH-radicals, the C$_{4'}$-O$_{4'}$ bond undergoes strong fluctuations close to $t\approx 500$ fs. It takes another 500 fs that C$_{4'}$-O$_{4'}$ bond breaks and the carbonyl-group forms. The establishment of carbonyl-group and its dynamical stability beyond $t\approx 1000$ fs is clearly visible. The C$_{4'}$-O$_{4'}$ distance in open-ring fluctuates around $d_{\rm CO}\approx 1.8 \AA$ due to thermal vibrations. For comparison, the time-evolution of C$_{4'}$-O$_{4'}$ bond distance corresponding to a deoxyribose far from OH-radicals is shown (dashed lines). It shows an undamaged bond length of $d_{\rm CO}\approx1.4\AA$, predicted by ReaxFF.
}
\label{Fig4d}
\end{center}\vspace{-0.5cm}
\end{figure}

\begin{figure}
\begin{center}
\includegraphics[width=1.1\linewidth]{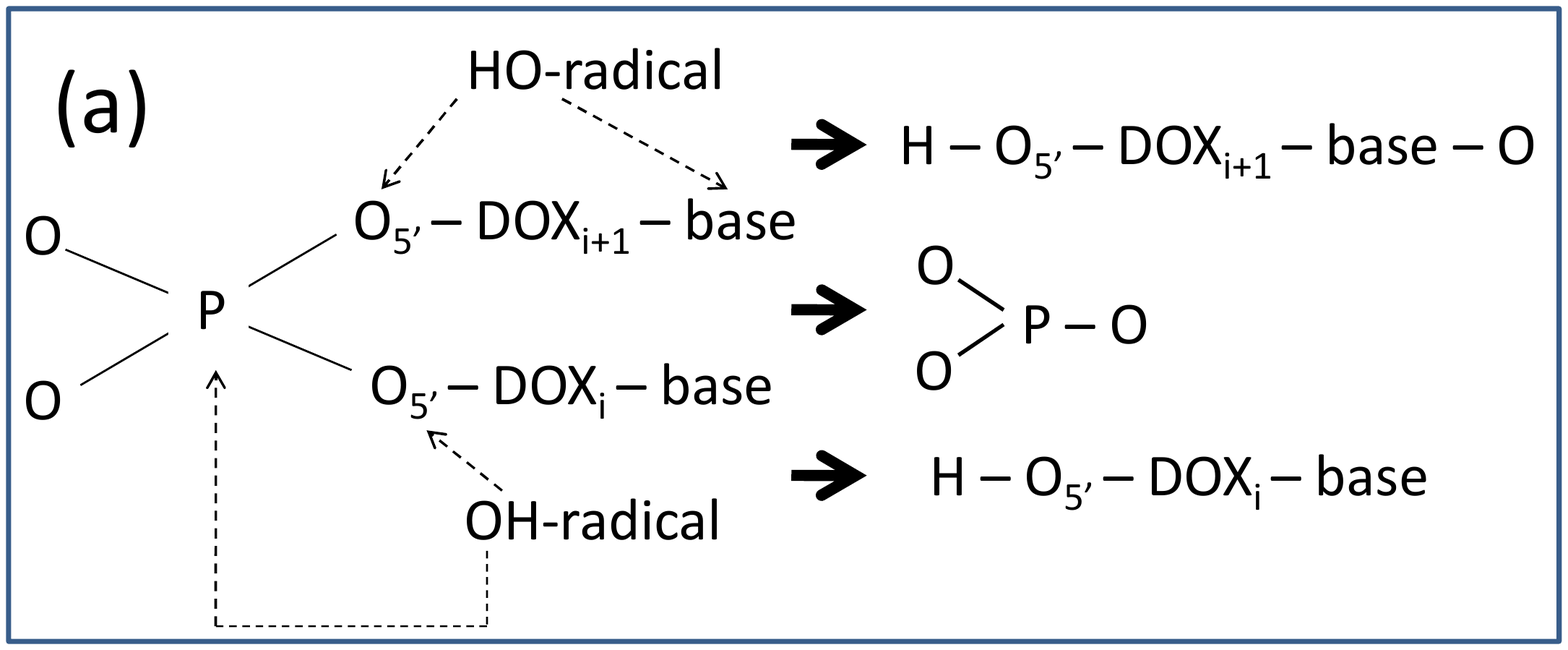}\\ \vspace{-3.5cm}
\includegraphics[width=1.0\linewidth]{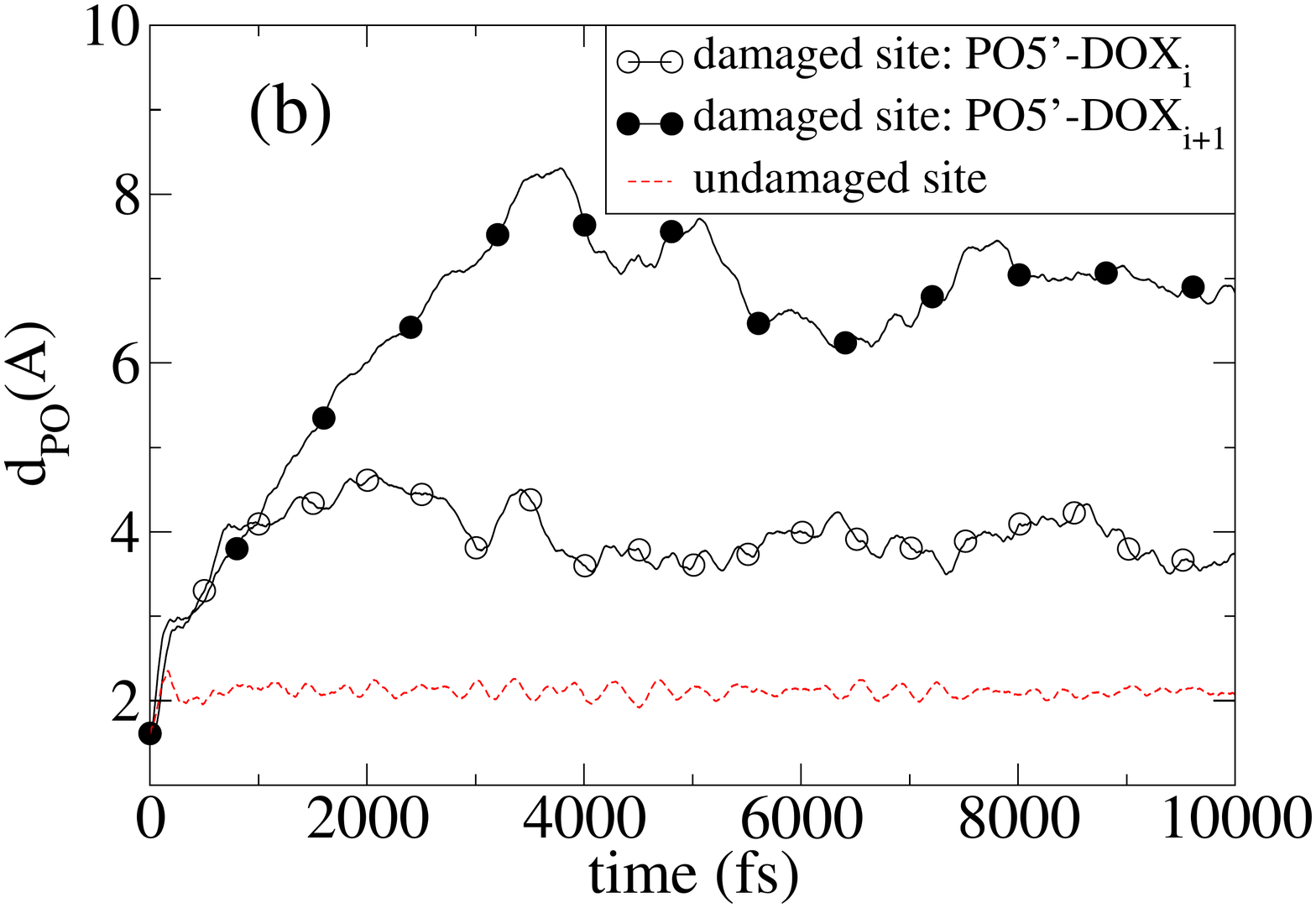}\\ 
\noindent
\caption{
(a) Schematic diagram of the chemical pathway leading to single strand break observed in ReaxFF MD. DOX$_i$ and DOX$_{i+1}$ denote two neighboring deoxyribose pairs labeled by the indices $i$th and $i+1$th. Donation of oxygen from one OH-radical lead to formation of free PO$_3$ and two disconnected DOX's.
(b) P-O$_{5'}$DOX$_i$ and P-O$_{5'}$DOX$_{i+1}$ bond distances as a function of time shown by circles, selected from a site of damage in which a cluster of OH-radicals are chemically active. Simultaneous increase in bond distances is a signature of the single strand break. For comparison the red dashed line shows typical P-O$_{5'}$ bond distance for a part of DNA back-bone that is far from OH-radicals.
}
\label{Fig4dd}
\end{center}\vspace{-0.5cm}
\end{figure}

Our model for the initial distribution of OH-radicals surrounding DNA molecule is based on well established description of water and ionizing-radiation interaction in which a mega-voltage beam interacts with water molecules through Compton effect and electron-positron pair production and produce spurs and blobs identified as small pockets of ion-pairs with typical diameter of 4-7 nm that fit approximately 3-12 ion pairs~\cite{EricHall:book}.

The information on the type of chemical reactions and the time-evolution of the damage is collected via running the MD up to 30 ps where the rearrangement of the atomic coordinates have been deduced from a dynamical trajectory calculated by ReaxFF. These simulations performed using periodic boundary conditions in a canonical moles, pressure and temperature (NPT) ensemble with a Nose-Hoover thermostat for temperature control and a time step of 0.25 fs.

Fig.~\ref{Fig1}(a) shows the initial structure of the DNA-molecule used in our simulation. The computational box comprise of DNA taken from the protein data base~\cite{PDB}, water molecules and randomly generated OH-radicals in small clusters around the DNA molecule.
The DNA molecule and solvated system was first energy minimized so as to eliminate any bad contact with water molecules arising from solvation. Small clusters of OH-radicals close to the DNA-backbone are shown within the circles.
Fig.~\ref{Fig1}(b) shows an early stage of the molecular-structure at $t=0.24$ ps, where the initial hydrogen-abstraction from DNA-backbone takes place. As it is seen, one OH-radical diffuses the distance $\ell\approx 6 \AA$ and removes $H_{5'}$ from the sugar-moiety, consistent with the experimental data~\cite{Pogozelski1998:CR,Tullius2005:CO} and CPMD calculation~\cite{Abolfath2009:JPC}.
Another interesting feature revealed in this simulation consists of the correlated states formed by OH-radicals that lead to their passivation. For example in Fig.~\ref{Fig1}(b) we observe that the two OH-radicals that do not participate in hydrogen-abstraction form hydrogen-bonds with smaller thermal diffusion length than an isolated OH-radical. In a cluster of OH-radicals, we find that not all of the OH-radicals thermally diffuse and interact with DNA.
Fig.~\ref{Fig2} reveals more details on $H_2O_2$ formation and their lower reactivity with DNA-molecule.
Consistent with our recent QM/MM calculation in the solution~\cite{Abolfath2011:JCP}, we observe that because of the hydrogen bond network forming between OH-radicals and water molecules the time for hydrogen abstraction is longer compare to similar simulation in vacuum.

\begin{figure}
\begin{center}
\includegraphics[width=1.0\linewidth]{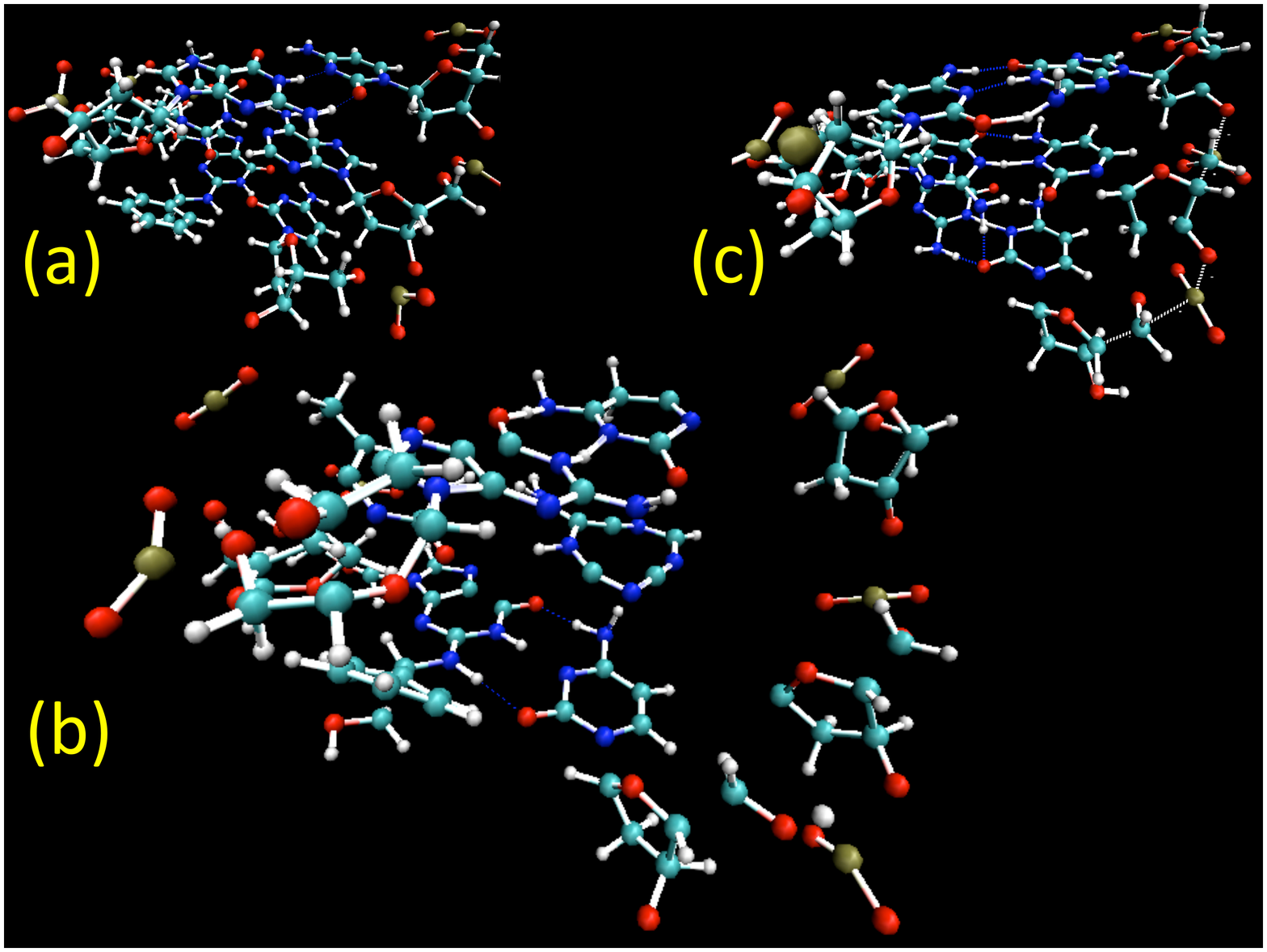}\\ 
\noindent
\caption{
(a) Initial configuration of a DNA-fragment in the presence of OH-radicals (not shown in the figure) and (b) the configuration of that part at $t=$2000 fs with a pronounced separated between bases and backbone. (c) For comparison another part of the DNA that has no exposure with OH-radicals is shown at $t=$2000 fs.
Carbon, oxygen, nitrogen, phosphorus and hydrogen atoms are shown as green, red, blue, gold and white, respectively.
The blue dash lines show the inter-base and base-stacking hydrogen bounds.
}
\label{Fig5}
\end{center}\vspace{-0.5cm}
\end{figure}

From ReaxFF-MD we realized that the initial damage that evolves to the separation of the nucleotide-base from DNA-backbone and DNA single strand break can be attributed to the formation of the hydroxyl- and carbonyl-groups in the sugar-rings in the backbone.
The oxygen from OH radical weaken the C-N bond that attaches the sugar-moiety with the nucleotide-base.
To investigate such possibility and the consistency of ReaxFF with DFT, we employed an ab-initio CPMD calculation in vacuum as well as a QMMM CPMD-GROMACS that allows adding realistic solution~\cite{Abolfath2011:JCP}.
The results are shown in Fig.~\ref{Fig3}. The molecular structure consists of deoxyribose sugar-moiety attached to amino group in the presence of OH-radical.
Fig.~\ref{Fig3} (d) and (e) show the result of CPMD in vacuum on hydroxyl- and carbonyl-formation and breaking of the bond in the sugar-ring that connects the amino-group to the backbone.

Our CPMD consists of four stages of wave function optimization, dynamical equilibration at T=300 K with ionic temperature control that allows step-wise increase of temperature, requenching of the wave function, and finally the microcanonical dynamics (constant energy ensemble) as described in Ref.~\cite{Abolfath2009:JPC}.
We have performed a constant pressure and temperature MD simulation with a reference temperature of 300K and a time step of 1fs. Our CPMD wave-function optimization is implemented in a plane-wave basis within local spin density approximation (LSDA) with an energy cutoff of 75 Rydberg (Ry), and with Becke~\cite{Becke1988:PRA} exchange and Lee-Yang-Parr (BLYP) gradient-corrected functional~\cite{Lee1988:PRB}. Norm conserving ultrasoft Vanderbilt pseudo-potentials were used for oxygen, hydrogen, nitrogen and carbon.
A cubic cell of size {$(13 \AA \times 13 \AA \times 13 \AA)$} is used together with Poisson solver of Martyna and Tuckerman~\cite{Martyna1999:JCP} for the wave function minimization in CPMD. For the QMMM calculation, the sugar-moiety, guanine (as a representative of a nucleotide base) and OH-radicals are in the QM part and the rest of DNA fragment including phosphorous and its two oxygen (O$_{\rm 1P}$ and O$_{\rm 2P}$) and water molecules are in MM part. The CPMD structural energy minimization and GROMACS force-field are implemented for QM and MM parts. The details of this calculation will be presented elsewhere~\cite{Abolfath2011:JCP}.

Fig.~\ref{Fig4} focuses on a fragment of the DNA that the base-sugar-moiety bond-breaking by OH-radicals is seen in ReaxFF. The initial configuration of the base (an adenine) is shown in Fig.~\ref{Fig4}(a). There are two OH-radicals close to the sugar-moiety (not shown in the figure). Two intermediate configurations are shown in Fig.~\ref{Fig4}(b) and (c) corresponding to $t=500$ fs and $t=3500$ fs respectively. Fig.~\ref{Fig4}(b) shows the damaged adenine and an undamaged cytosine that is one base below the adenine (far from OH-radicals). In Fig.~\ref{Fig4}(c) we show the damaged adenine with another cytosine (one base above the adenine). Both adenine and cytosine were initially close to free radicals.

A substantial damage in adenine-groups into a ring and chain is clearly observed. While some of this distortion may be related to secondary reactions related to the OH-damage, it is possible that the current ReaxFF description overestimates the distortion rate. We aim to replicate the distortion pathway at the DFT-level to confirm the results obtained by ReaxFF and if necessary will improve this aspect of the ReaxFF description to make it consistent with DFT. We note that based on our experimental results, such a substantial distortion is qualitatively consistent with the absorption spectrum of unirradiated and that receiving ionizing radiation where the effect of irradiation can be visible in the blurring of the absorbance peaks. Our  quantitative comparison between computational modeling and experimental data obtained from photo-emission is on the way~\cite{AbolfathKodym}.
Finally the observed damage in cytosine is mainly associated with a dislocated hydrogen. As it is shown in the figure, both bases are moved away from sugar-moiety with a distance that is approximately around 5 $\AA$, hence they are completely disconnected from the backbone.

In Fig.~\ref{Fig4}(c) we observe an opened gap in deoxyribose-ring due to C$_{4'}$-O$_{4'}$ broken bond. The corresponding time evolution of the bond-distance is shown in Fig.~\ref{Fig4d}. The bond undergoes a severe fluctuation in $t\approx 500-1000$ fs due to occurrence of series of chemical reactions by two OH-radicals initially located at opposite sides of deoxyribose-ring including (a) hydrogen abstraction of H$_{3'}$ by OH radical  incorporated by formation of water molecule and (b) formation of hydroxyl group with C$_{4'}$. Finally beyond $t\approx 1000$ fs the open-gap with
$d_{\rm CO}\approx 1.8\AA$ is established. These collection of events including base-damage incorporated with formation of a hydroxyl- and carbonyl-group is consistent with the picture deduced by ab-initio CPMD and QMMM.

We now turn to DNA back-bone damage and observation of the single strand break (SSB) in ReaxFF MD. We define $d_{\rm PO}$ as the relative distance between P and two neighboring deoxyribose-rings denoted by DOX$_i$ and DOX$_{i+1}$ as shown in Fig.~\ref{Fig4dd}(a). This figure summarizes a pathway that leads to SSB observed in ReaxFF MD. Two OH-radicals participate in the chemical reaction. One OH donates an oxygen to phosphorous in the back-bone that results to formation of PO$_3$ and strong fluctuation in the links between P and two neighboring DOX that finally lead to disjointing two DOX's from back-bone as shown in Fig.~\ref{Fig4dd}(b).
Here solid lines marked by circles show the time evolution of P-DOX$_i$ and P-DOX$_{i+1}$ bond-distances close to one of the clusters of OH-radicals.
Note that in our simulation SSB has been observed only in locations where OH-radicals actively participate in chemical reactions.
Clearly simultaneous stretch in P-DOX$_i$ and P-DOX$_{i+1}$ bonds is indication of concerted events that lead to SSB. For comparison a typical time evolution of $d_{\rm PO}$ is shown (the dashed-line) selected from an undamaged site of DNA located far from OH-radicals. As it is seen, ReaxFF predicts an average bond-length of $d_{\rm PO}\approx 2 \AA$. Note that the cut-off, set for bond visualization, chosen for carbon bonds with $d=1.5\AA$. The phosphate fragments seen in the figures are the artifact of VMD visualization.
The remaining H from the fragmented OH-radical either forms a water molecule by combining with the other OH-radical [not shown in Fig.~\ref{Fig4dd}(a)] or passivate the dangling bond of O$_{5'}$-DOX$_i$ and forming a hydroxyl group. In the latter, the other OH disintegrates to H and O where H passivates the dangling bond of O$_{5'}$-DOX$_{i+1}$  and O oxidizes the base [see Fig.~\ref{Fig4dd} (a)].

Fig.~\ref{Fig5} reveals similar features in a larger scale. Here the initial and final configurations corresponding to $t=2$ ps are shown in Fig.~\ref{Fig5}(a) and (b). For comparison, a part of DNA that is far from concentration of OH-radicals is shown in Fig.~\ref{Fig5}(c). The damage and disconnectivity of the bases-backbone is clearly visible in Fig.~\ref{Fig5}(b), where the stability of the DNA in the absence of OH-radicals is obvious from Fig.~\ref{Fig5}(c). The horizental/vertical blue dash-lines show the base pair/stacking hydrogen bonds. As shown in Fig.~\ref{Fig5}(b) the network of hydrogen bonds is disrupted because of the damaged sites. In contrast, in undamaged fragment, as shown in  Fig.~\ref{Fig5}(c), the network of hydrogen bonds are preserved by MD.

In conclusion, we have examined the MD simulation based on ReaxFF on a large fragment of DNA molecule to simulate the hydrogen abstraction by OH-radicals. The simulation reveals various type of dehydrogenation from DNA back-bone and DNA-base that evolve into the disconnectivity of the bases from the backbone due to carbonyl and hydroxyl formation in sugar-moiety by OH-radicals. We further confirmed the results obtained by ReaxFF-MD using an ab-initio CPMD calculation. Close examination of damage with various length scales show that the damage collectively disrupt the base-pair hydrogen bonds and lead to base-backbone bond breaking, single-strand break and finally double strand break.
The present work that is in complement with previous calculation on smaller system in vacuum and solution using DFT and QM/MM can be used as a computational platform for energy scoring and the effects of radiation on biological systems.

Authors would like to thank Dr. Kyeongjae Cho and Dr. Yves Chabal
for useful comments and discussion.




\end{document}